\documentclass[12pt]{iopart}

\usepackage{epsfig}

\newcommand{\ltsim}{\,{\buildrel < \over {_\sim}}\,}
\newcommand{\gtsim}{\,{\buildrel > \over {_\sim}}\,}

\newcommand{\commented}[1]{{}}

\begin{document}

\title[]{Enhanced charm hadroproduction due to nonlinear corrections
to the DGLAP equations\footnote[7]{The work of K.J.E. and V.J.K. was supported
by the Academy of Finland, projects 50338, 80385 and 206024.
The work of R.V. was supported in part by the Director, Office of
Energy Research, Division of Nuclear Physics of the Office of High
Energy and Nuclear Physics of the U. S.  Department of Energy under
Contract Number DE-AC03-76SF00098. }}

\author{K.J. Eskola$^{a,b}$, 
\underline{V.J. Kolhinen}$^{a,b}$\footnote[1]{email: vesa.kolhinen@phys.jyu.fi}
and R. Vogt$^{c,d}$}

\address{$^a$ Department of Physics, 
P.O.B. 35, FIN-40014 University of Jyv\"askyl\"a, Finland}

\address{$^b$ Helsinki Institute of Physics,
P.O.B. 64, FIN-00014 University of Helsinki, Finland}

\address{$^c$ Lawrence Berkeley National Laboratory, Berkeley, CA 94720, USA}

\address{$^d$ Physics Department, University of California, Davis, CA 95616, USA }

\begin{abstract}

We have studied the effects of nonlinear scale evolution of the parton
distribution functions to charm production in $pp$ collisions at
center-of-mass energies of 5.5, 8.8 and 14 TeV. We find that the
differential charm cross section can be enhanced up to a factor of 4-5
at low $p_T$.  The enhancement is quite sensitive to the charm quark
mass and the renormalization/factorization scales.

\end{abstract}

%Uncomment for PACS numbers title message
%\pacs{00.00, 20.00, 42.10}

% Uncomment for Submitted to journal title message
%\submitto{\JPA}

% Comment out if separate title page not required
% \maketitle

Global fits of parton distribution functions (PDFs) have been obtained
by several groups, such as CTEQ \cite{Pumplin:2002vw,Stump:2003yu} and
MRST \cite{Martin:2001es,Martin:2002dr,Martin:2003sk}. These fits,
based on the Dokshitzer-Gribov-Lipatov-Altarelli-Parisi (DGLAP)
\cite{DGLAP} scale evolution, fit the deep inelastic scattering (DIS)
HERA data \cite{Adloff:2000qk} on the proton structure function
$F_2(x,Q^2)$ at large interaction scales $Q^2 \gtsim 10$ GeV$^2$ and
momentum fractions $x \gtsim 0.005$ very well. However, at small
scales, $Q^2 \ltsim 4$ GeV$^2$, and at small momentum fractions, $x
\ltsim 0.005$, these fits are usually worse.  In addition, the
next-to-leading order (NLO) gluon distributions become very small or
even negative in the small-$x$, small-$Q^2$ region. In this region,
the gluon recombination terms, giving rise to nonlinear corrections to
the evolution equations, become important. In a previous work
\cite{Eskola:2002yc}, the first of these nonlinear terms, calculated
by Gribov, Levin and Ryskin (GLR) \cite{GLR} and Mueller and Qiu (MQ)
\cite{Mueller:wy}, were included in the leading order (LO) DGLAP
evolution equations of gluons and sea quarks.  It was shown that the
HERA DIS $F_2(x,Q^2)$ data can be reproduced well with a new PDF
set\footnote{Available at www.urhic.phys.jyu.fi.}, EHKQS, obtained in
Ref.~\cite{Eskola:2002yc}, employing LO-DGLAP+GLRMQ evolution.

\begin{figure}
\includegraphics[width=8cm]{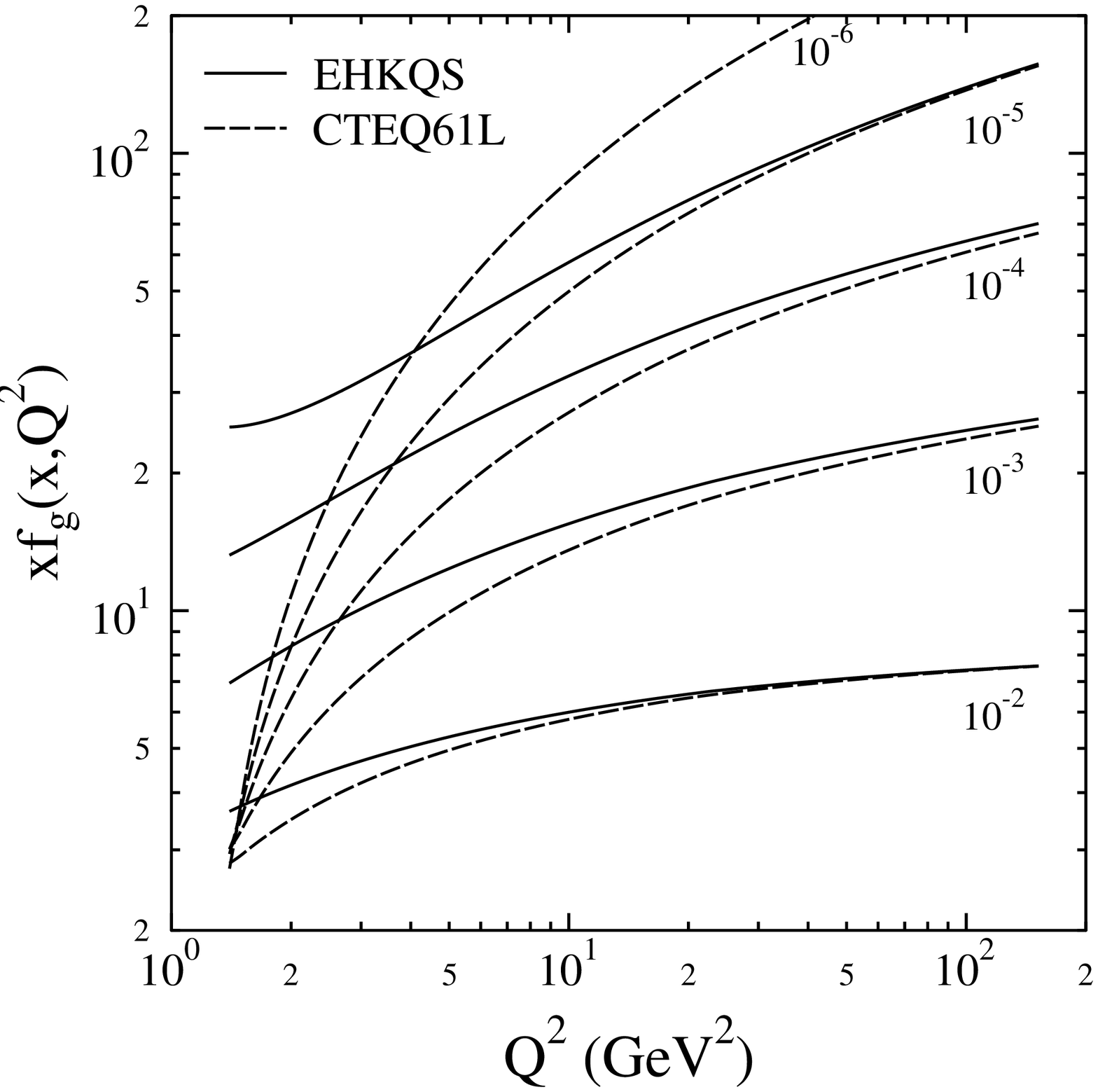}
\includegraphics[width=8cm]{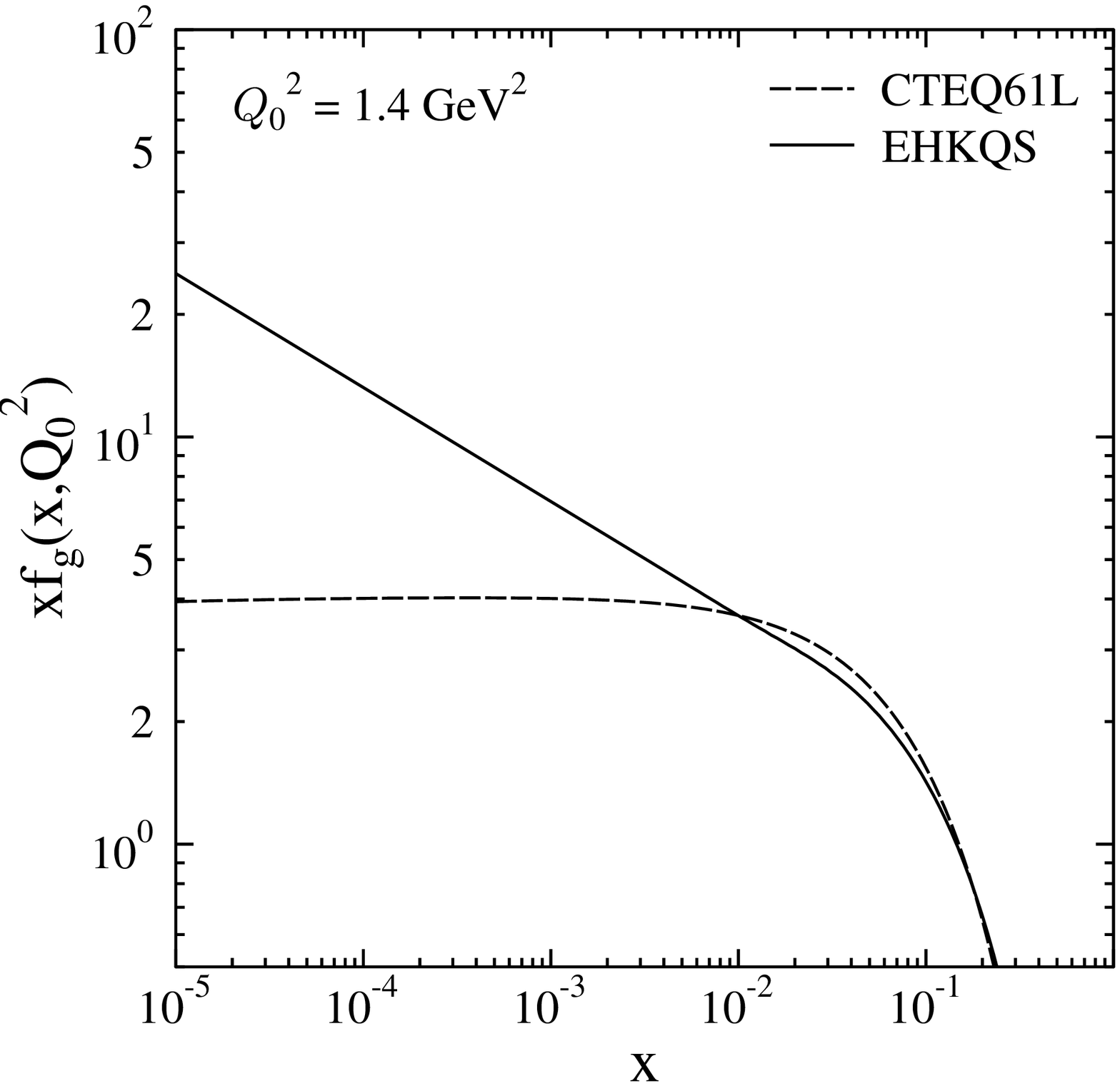}
\caption[]{\small {\bf Left:} Scale evolution of the EHKQS and CTEQ61L
gluon distribution for various fixed values of $x$. The enhancement
caused by the nonlinear terms vanishes during the evolution.  {\bf
Right:} The gluon distribution as a function of $x$ at $Q^2=1.4$
GeV$^2$. }
\label{FigGluon}
\end{figure}

Introducing the GLRMQ terms slows the scale evolution. At $Q^2 \ltsim
10$ GeV$^2$ and $x \ltsim 0.01$, where the nonlinearities are
important, this gives rise to larger gluon distributions than in the
pure DGLAP case.  This is shown on the left-hand side of
Fig.~\ref{FigGluon}, where we plot the scale evolution of the EHKQS
and CTEQ61L \cite{Stump:2003yu} PDF sets for several fixed values of
$x$. The enhancement, large at small scales, vanishes as the nonlinear
terms become negligible at larger scales. Finally, at $Q^2 \gtsim 10$
GeV$^2$, the evolution is clearly dominated by the DGLAP terms.  On
the right-hand side of Fig. \ref{FigGluon} we show the gluon
distributions as a function of $x$ for $Q^2=1.4$ GeV$^2$, the initial
scale of the EHKQS set. As seen in Fig.~\ref{FigGluon}, the HERA data
relevant for nonlinear scale evolution suggests a factor of $\sim 3$
gluon enhancement at $x\sim 10^{-4}$ and $Q^2=1.4$ GeV$^2$.

However, since the HERA $F_2$ data alone cannot distinguish between
the linear DGLAP and nonlinear DGLAP+GLRMQ evolution, additional
independent probes are needed.  Here, we discuss how charm quark
production in $pp$ collisions could probe the gluon enhancement
\cite{Eskola:2003fk}.  Charm production is an ideal choice since the
charm mass is low and its production is dominated by gluons. Assuming
factorization, inclusive differential charm cross sections at high
energies can be expressed as
\begin{equation}
 d\sigma_{pp\rightarrow c \overline c X}(Q^2,\sqrt s) = \hspace{-0.5cm}
   \sum_{i,j,k=q,\overline q,g} \hspace{-0.4cm}
   f_i(x_1,Q^2)\otimes f_j(x_2,Q^2)
   \otimes d\hat \sigma_{ij\rightarrow c \overline c \{k\}}(Q^2,x_1,x_2)
\label{sigcc}
\end{equation}
where $\hat \sigma_{ij\rightarrow c \overline c \{k\}}(Q^2,x_1,x_2)$
are the perturbatively calculable partonic cross sections for charm
production at scales $Q^2 \sim m_T^2 \gg\Lambda^2_{\rm QCD}$, $x_1$
and $x_2$ are the momentum fractions of the partons involved in the
hard scattering and $f_i(x,Q^2)$ are the free proton PDFs. We assume
that the renormalization scale and factorization scale are equal.
Only the $gg$ and $q \overline q$ channels are allowed at LO, which we
consider here.

\begin{figure}
\center{\includegraphics[width=13cm]{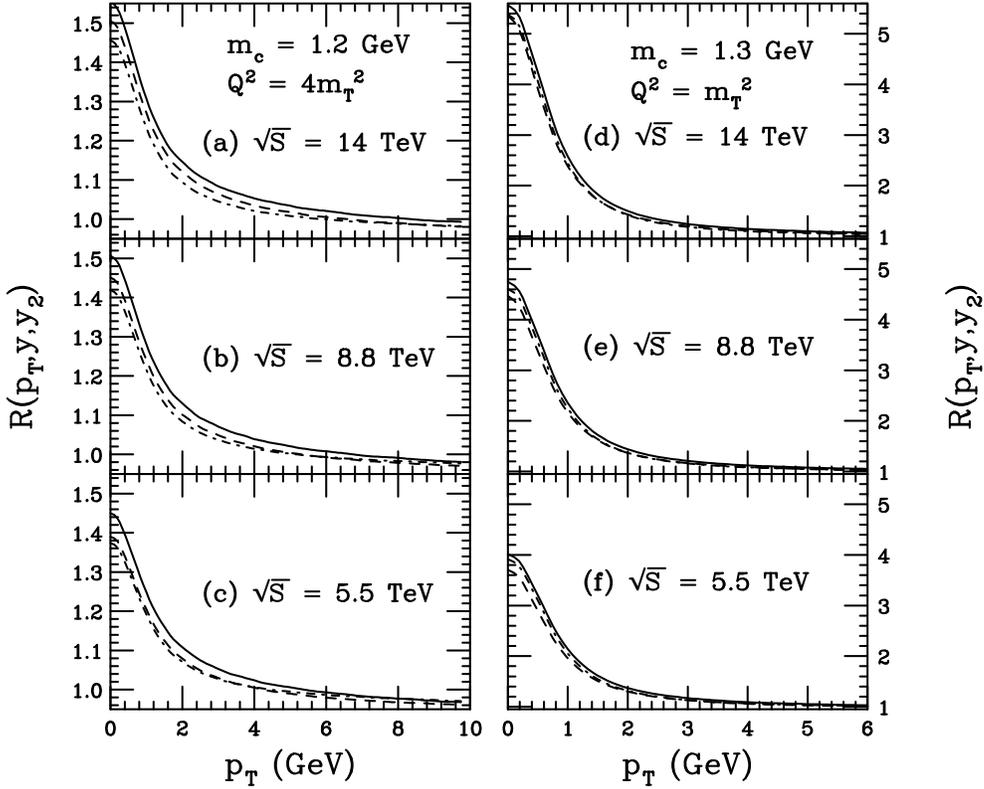}}
\caption[]{\small The ratio of differential charn cross section,
$R(p_T,y,y_2)$, for fixed $y$ and $y_2$ as a function of charm quark
$p_T$ at $\sqrt{s}=14$ TeV (top) 8.8 TeV (middle) and
5.5 TeV (bottom) in $pp$ collision.  The rapidities are $y=y_2=0$
(solid), $y=2$, $y_2=0$ (dashed) and $y=y_2=2$ (dot-dashed).  }
\label{FigR}
\end{figure}

We calculate the ratio of differential cross sections, 
\begin{equation}
  R(p_T,y,y_2) \equiv 
    \frac{d^3\sigma({\scriptstyle\rm EHKQS})/(dp_T dy dy_2)}
         {d^3\sigma({\scriptstyle\rm CTEQ61L})/(dp_T dy dy_2)},
\label{Rdsigmas}
\end{equation}
where $p_T$ is the charm quark transverse momentum and $y$ and $y_2$
are the rapidities of the charm quark and the antiquark. The results
for the enhancement are plotted in Fig.~\ref{FigR} as a function of
$p_T$ for fixed $y$ and $y_2$.  The center-of-mass energy $\sqrt{s}$
is varied from 14 TeV (top) to 8.8 TeV (middle) and 5.5 TeV
(bottom). Obviously, the largest enhancement is obtained at the
largest $\sqrt{s}$ where the $x$ values are smallest.  Different charm
masses and scales are used, $m_c=1.2$ GeV and $Q^2=4m_T^2$ on the
left-hand side and $m_c=1.3$ GeV, $Q^2=m_T^2$ on the right-hand side
\cite{Vogt:2001nh}. Both of these scale choices lie within the
applicability region of the PDFs.  The rapidities are $y=y_2=0$
(solid), $y=2$, $y_2=0$ (dashed) and $y=y_2=2$ (dot-dashed). For the
highest energy, 14 TeV, the maximum $\sqrt{s}$ at the LHC, the
enhancement at $p_T\sim 0$ is a factor of $\sim 5$ for $m_c=1.3$ GeV
and $\sim 1.5$ for $m_c=1.2$ GeV.  We repeated the calculations for
larger masses, up to $m_c=1.8$ GeV for both $Q^2=m_T^2$ and
$Q^2=4m_T^2$.  We found smaller enhancements, $\sim$ 2 and $\sim$ 1.25
at $p_T\sim 0$, respectively.  Clearly, the charm enhancement can be
substantial, but it is very sensitive to the choice of mass and scale.
It also vanishes rapidly with $p_T$.

Integrating over the rapidities does not change the result much. The
maximum enhancement for the rapidity-integrated ratios at small $p_T$
at $\sqrt{s}=14$ TeV is still $\sim 4.5$ for $m_c=1.2$ GeV,
$Q^2=4m_T^2$ and $\sim 1.3$ for $m_c=1.3$ GeV, $Q^2=m_T^2$
\cite{Eskola:2002yc}.  The ratio of the cross sections integrated over
$p_T$ and $y_2$ is rather flat as a function of $y$, but is reduced to
factors of 1.2 and 1.8 for the two low mass cases studied.

Since the DGLAP gluon distributions are already well constrained by
the HERA data, they cannot absorb additional large effects.  Therefore
we can conclude that, if this small-$p_T$ enhancement in the charm
cross section relative to the DGLAP-based result is observed in the
future experiments e.g. at the LHC, it is a signal of nonlinear
effects on the PDF evolution.

As discussed in Ref. \cite{Eskola:2002yc}, the EHKQS PDFs were
obtained in the region where the nonlinear terms do not dominate the
scale evolution. However, at the smallest $x$ and $Q^2$ of the HERA
data \cite{Adloff:2000qk} one is already close to the gluon saturation
region, where all orders of nonlinearities become important
\cite{Eskola:2003gc}. Therefore these PDFs, and consequently also the
charm enhancement, should be considered as upper limits.  Furthermore,
a full NLO DGLAP+GLRMQ analysis would be needed for computing the
charm hadroproduction enhancement consistently to NLO.

In a follow-up work \cite{Dainese}, we show that more than half of the
charm enhancement survives hadronization to $D$-mesons.  We also show
that in the most optimistic case the enhancement can be observed in
the $D^0$ $p_T$ spectrum in the ALICE detector at the LHC.

\end{document}